\documentclass[a4paper,12pt]{article}

\usepackage{xspace}
\usepackage{amsmath}
\usepackage{natbib}
\usepackage{comment}
\usepackage{graphicx}
\usepackage{authblk}
\usepackage[colorlinks=true,citecolor=blue]{hyperref}

\newcommand{\ave}[1]{\left\langle#1 \right\rangle}

\newcommand{\elabel}[1]{\label{eq:#1}}
\newcommand{\eref}[1]{(Eq.~\ref{eq:#1})}

\newcommand{\Eref}[1]{Equation~(\ref{eq:#1})}
\newcommand{\flabel}[1]{\label{fig:#1}}
\newcommand{\fref}[1]{Fig.~\ref{fig:#1}}

\newcommand{\seclabel}[1]{\label{sec:#1}}
\newcommand{\secref}[1]{Sec.~\ref{sec:#1}}

\newcommand{\gtime}{\bar{g}}

\newcommand{\Ito}{It\^{o}\xspace}

\newcommand{\ie}{{\it i.e.}\xspace}

\newcommand{\cf}{{\it c.f.}\xspace}

\newcommand{\be}{\begin{equation}}
\newcommand{\ee}{\end{equation}}
\newcommand{\bea}{\begin{eqnarray}}
\newcommand{\eea}{\end{eqnarray}}
\newcommand{\D}{\Delta}
\newcommand{\yn}[1]{{y^{(#1)}}}
\newcommand{\xin}[1]{{\xi^{(#1)}}}
\newcommand{\Wn}[1]{{W^{(#1)}}}
\newcommand{\mun}[1]{{\mu^{(#1)}}}
\newcommand{\sigman}[1]{{\sigma^{(#1)}}}
\newcommand{\sigmac}{{\sigma_\rho}}

\title{An evolutionary advantage of cooperation}
\author{
Ole Peters$^{1,2,\dagger}$ and Alexander Adamou$^{1,\ddagger}$\\
\normalsize
$^1$London Mathematical Laboratory, 8 Margravine Gardens, London W6 8RH, UK\\
$^2$Santa Fe Institute, 1399 Hyde Park Road, Santa Fe, 87501 NM, USA\\
$^\dagger$\texttt{o.peters@lml.org.uk}, $^\ddagger$\texttt{a.adamou@lml.org.uk}
}
\date{May 24, 2018}

\begin{document}
\maketitle

\begin{abstract}
Cooperation is a persistent behavioral pattern of entities pooling and sharing resources. Its ubiquity in nature poses a conundrum. Whenever two entities cooperate, one must willingly relinquish something of value to the other. Why is this apparent altruism favored in evolution? Classical solutions assume a net fitness gain in a cooperative transaction which, through reciprocity or relatedness, finds its way back from recipient to donor. We seek the source of this fitness gain. Our analysis rests on the insight that evolutionary processes are typically multiplicative and noisy. Fluctuations have a net negative effect on the long-time growth rate of resources but no effect on the growth rate of their expectation value. This is an example of non-ergodicity. By reducing the amplitude of fluctuations, pooling and sharing increases the long-time growth rate for cooperating entities, meaning that cooperators outgrow similar non-cooperators. We identify this increase in growth rate as the net fitness gain, consistent with the concept of geometric mean fitness in the biological literature. This constitutes a fundamental mechanism for the evolution of cooperation. Its minimal assumptions make it a candidate explanation of cooperation in settings too simple for other fitness gains, such as emergent function and specialization, to be probable. One such example is the transition from single cells to early multicellular life.
\end{abstract}


\newpage
\section{Introduction}
\seclabel{intro}
\begin{center}
{\it They give that they may live, for to withhold is to perish.}\\
K. Gibran
\end{center}

\citet[p.~1563]{Nowak2006} concludes his review of the mechanisms of cooperation with the words: ``Perhaps the most remarkable aspect of evolution is its ability to generate 
cooperation in a competitive world.'' Indeed, life is full of cooperative structure. We living beings exist not as minimal self-reproducing chemical units, but as cells, organisms, families, herds, companies, institutions, nations, and so on. Cooperation -- the persistent behavioral pattern of entities pooling and sharing their resources -- is ubiquitous in nature.

This ubiquity is puzzling because pooling and sharing seem \textit{prima facie} to involve altruism. The temporarily better-off entity in a cooperating pair must willingly relinquish something of value to the worse-off entity to maintain the cooperative pact. If naked altruism is an unsatisfactory explanation of evolved behavior, then we must elucidate the benefit derived by the better-off entity in such an arrangement.

Classical explanations involve two ideas. The first is that a cooperation benefit exists between two entities. Specifically, the fitness gain of the recipient exceeds the fitness cost to the donor. The second is that, over time, this net increase in fitness finds its way back to the donor. This can happen through reciprocity, where past donors become future recipients, or through relatedness, where the recipient carries genetic material that the donor is interested in propagating. \citet{Nowak2006} offers a comprehensive account of this approach, delineating five cooperative arrangements possible when a cooperation benefit exists, and five corresponding conditions under which cooperation is favored in evolution.

Our aim here is to postulate a mechanism through which this cooperation benefit, or net fitness gain, arises. This is important because classical analyses are predicated on its existence.

We start from the basic model of a living entity as something which self-reproduces with temporal fluctuations. Treating biomass as a multiplicative stochastic process yields two exponential growth rates: that achieved in the long-time limit, on which fluctuations have a negative effect; and that achieved in the many-entities limit, unaffected by fluctuations. This is a manifestation of non-ergodicity.

We hypothesize that repeated pooling and sharing is beneficial because, by reducing the net effect of fluctuations, it increases the long-time growth rate at which cooperating entities self-reproduce. Cooperation is observed in nature because those who do it outgrow those who don't. We identify this increase in the growth rate of biomass with the net fitness gain of classical treatments.

Fitness has many definitions in the biological literature \citep{Orr2009}. While commonly agreed to refer to the ability of living organisms to survive and reproduce, no precise definition of it has achieved consensus. Of all the definitions proposed, our work relates most closely to geometric mean fitness, on which the effects of fluctuations are recognized \citep{LewontinCohen1969,Gillespie1977,Orr2009}. To avoid confusion, we speak here of growth rates rather than fitness, since they are defined unequivocally. Among entities undergoing noisy multiplicative growth, that whose time-average growth rate is highest will, over time, come to dominate its environment.

Motivating this study is a particular evolutionary phenomenon: the transition to multicellularity. This occurs when a species of non-cooperating single cells mutates to a species of multicellular organisms, sharing nutrients through common membranes. Often this is explained by a different type of fitness gain: the emergence of new function. An agglomeration of cells may, for example, develop the ability to swim up a nutrient gradient or funnel resources towards itself \citep{ShortETAL2006,RoperETAL2013}. Implicit in such explanations is a degree of complexity not present in early unicellular life. The performance of specialized tasks requires many cells to be assembled in delicate designs, whose spontaneous development would be extremely improbable. To be a credible theory, evolution must explain not only the rich tapestry of cooperative structure we observe now, but also early cooperative steps, such as from single cell to cell pair, where new function cannot be relied upon. A universal mechanism for the evolutionary advantage of cooperation is needed.

We present our work as follows. In \secref{growth} we introduce a simple mathematical model of noisy multiplicative growth and summarize its relevant properties. In \secref{coop} we describe a cooperation protocol in which entities grow, pool, and share biomass. In \secref{analysis} we show that, under certain conditions, entities which cooperate increase the time-average growth rate of their biomass. This, we hypothesize, is a universal explanation of the existence of cooperation. In \secref{general} we discuss generalizations of our model. We offer concluding remarks in \secref{discussion}.

\section{Noisy multiplicative growth}
\seclabel{growth}
Let $x_i(t)$ represent the biomass of an entity $i$ at time $t$. Biomass generates more of itself. For example, a cell collects nutrients and other matter, then splits into two cells, which split into four cells, and so on. This happens stochastically. Some entities thrive, perhaps in safe and plentiful environments, while others have their growth hampered, sometimes terminally.

While we speak typically of the biomass of an entity, our analysis is agnostic to replacement of `biomass' by `population,' `resources,' `wealth,' and the like; and of `entity' by `cell,' `organism,' `colony,' `herd,' `society,' and so on. Cooperation in nature occurs in many domains and at many scales.

A common and general model of noisy self-reproduction is geometric Brownian motion. In simple terms, the change in biomass over a short time step is a normally distributed random multiple of the existing biomass. More formally, $x_i(t)$ follows the \Ito drift-diffusion process,
\begin{equation}
dx_i=x_i(\mu dt + \sigma dW_i),
\elabel{motion}
\end{equation}
where $\mu$ is the drift and $\sigma$ is the volatility. The $dW_i$ are independent and identically-distributed random increments of the Wiener process, which are normal with zero mean and variance $dt$.

Geometric Brownian motion is a universal model because it is an attractor for more complex models that exhibit multiplicative growth \citep{AitchisonBrown1957,Redner1990}. It is a model of unconstrained growth. Self-reproduction limited by resource or space constraints or by predation would be poorly described by \eref{motion}. The water lily of the famous riddle, recounted in \citep{MeadowsETAL1972}, stops growing exponentially once it covers the pond.

Over time $T$, each biomass experiences a random exponential growth rate, defined as\begin{equation}
g(x_i,T) \equiv \frac{1}{T} \ln\left(\frac{x_i(T)}{x_i(0)}\right).
\elabel{g_indiv}
\end{equation}
Imagine starting many cell cultures in separate petri dishes and watching their biomasses evolve according to \eref{motion} for time $T$. Assume the dishes are large enough and $T$ short enough that growth does not slow for want of agar. The observed exponential growth rates would be drawn from a normal distribution with mean $\mu-\sigma^2/2$ and variance $\sigma^2/T$:
\be
g(x_i,T) \sim \mathcal{N}\left(\mu-\frac{\sigma^2}{2}, \frac{\sigma^2}{T}\right).
\elabel{g_dist}
\ee 

The expectation value or ensemble average of the biomass is defined as
\be
\ave{x} \equiv \lim_{N\rightarrow\infty} \frac{1}{N}\sum_{i=1}^N x_i.
\ee
Its evolution is computed by noting that \eref{motion} implies $d\ave{x}=\mu\ave{x}dt$. This describes exponential growth of $\ave{x}$ at a rate equal to the drift. We will call this the \textit{ensemble-average growth rate} and denote it by
\be
g(\ave{x})=\mu.
\elabel{gave}
\ee
The physical meaning of this quantity is worth making explicit. It is the growth rate of the large-sample limit of mean biomass. Since \eref{motion} is not an ergodic process, this observable is {\it a priori} uninformative of the temporal behavior of a given trajectory. We don't mean by this the trivial observation that individual trajectories are stochastic, while their expectation value is not. Rather, there are two fundamentally different ways of removing stochasticity from \eref{g_indiv}: we may compute \eref{gave}; or we may consider a single long trajectory, and let time remove randomness from the growth rate. 

If only a single system is to be modeled, then \eref{gave} is, in essence, fiction: the expectation value is an average over imagined parallel universes, where the randomness in \eref{motion} plays out in all its different possible ways.

One might guess that the growth rate observed in an individual trajectory will converge to \eref{gave} over time, but that is simply a common conceptual error. Instead, the non-ergodicity of \eref{motion} manifests itself such that the growth rate observed in a single trajectory converges to a different value, which we call the \textit{time-average growth rate}. This is the almost sure limit of \eref{g_dist} as $T\to\infty$,
\be
\gtime(x_i) \equiv \lim_{T\rightarrow\infty}g(x_i,T) = \mu-\frac{\sigma^2}{2}.
\elabel{gt}
\ee
We note that $\gtime(x_i)=\ave{g(x_i,T)}$, meaning that $g(x_i,T)$ is an ergodic observable (specifically, one whose expectation value reflects what happens over time in a single trajectory) for the non-ergodic process of noisy multiplicative growth \citep{PetersGell-Mann2016}.

We see in nature what has survived. In our model, the entity with the highest time-average growth rate will, regardless of its ensemble-average growth rate, come to dominate the biomass in the system over time. The ratio of its biomass to that of other entities will grow exponentially. Strategies which maximize $\gtime(x_i)$, and not necessarily $g(\ave{x})$, will confer an evolutionary advantage on their adherents. For any entity, $\gtime(x_i)$ is less than $g(\ave{x})$ by the fluctuation-sensitive term $\sigma^2/2$. This standard result of \Ito calculus \citep[Ch.~5]{vanKampen1992} has found its way into the biological literature \citep[Eq.~1]{Gillespie1977}. It suggests that strategies which reduce volatility should be favored in evolution and observed in nature.

The inequality of the ensemble-average and time-average growth rates in geometric Brownian motion, discussed in \citep{PetersGell-Mann2016, AdamouPeters2016} in the context of economics, is known by evolutionary biologists. A clear description is found in \citep[p.~1836]{Lande2007}. The difference is often labelled as that between arithmetic and geometric mean growth rates \citep{LewontinCohen1969} or between arithmetic and geometric mean fitnesses \citep{Orr2009}. The positive effect of volatility reduction on geometric mean fitness is also recognized, as noted in \citep{Gillespie1977} and \citep[Box~3]{Orr2009}. It is precisely this effect in the context of cooperation that we explore here.

\section{Cooperation protocol}
\seclabel{coop}
Having established these properties of \eref{motion}, we now introduce
our model of cooperation. Our reference point is a population of $N$ non-cooperating entities, such as single-celled organisms, whose biomasses follow geometric Brownian motions with identical drift and volatility but with independent realizations of the noise. In other words, \eref{motion} with $i=1 \dots N$.

Consider a discretized version of \eref{motion}, such as would be used in a numerical simulation. The non-cooperators grow according to
 \bea
 \D x_i(t) & = & x_i(t) \left(\mu \Delta t + \sigma \xi_i \sqrt{\D t}\right), \elabel{discrete_nonc_grow} \\
 x_i(t+\D t) & = & x_i(t) + \D x_i(t), \elabel{discrete_nonc_coop}
 \eea
where $\xi_i$ are independent standard normal variates, $\xi_i \sim \mathcal{N}(0,1)$.

The cooperation mechanism, summarized pictorially for $N=2$ in \fref{dynamics}, is as follows.
\begin{figure}
\includegraphics[width=\textwidth]{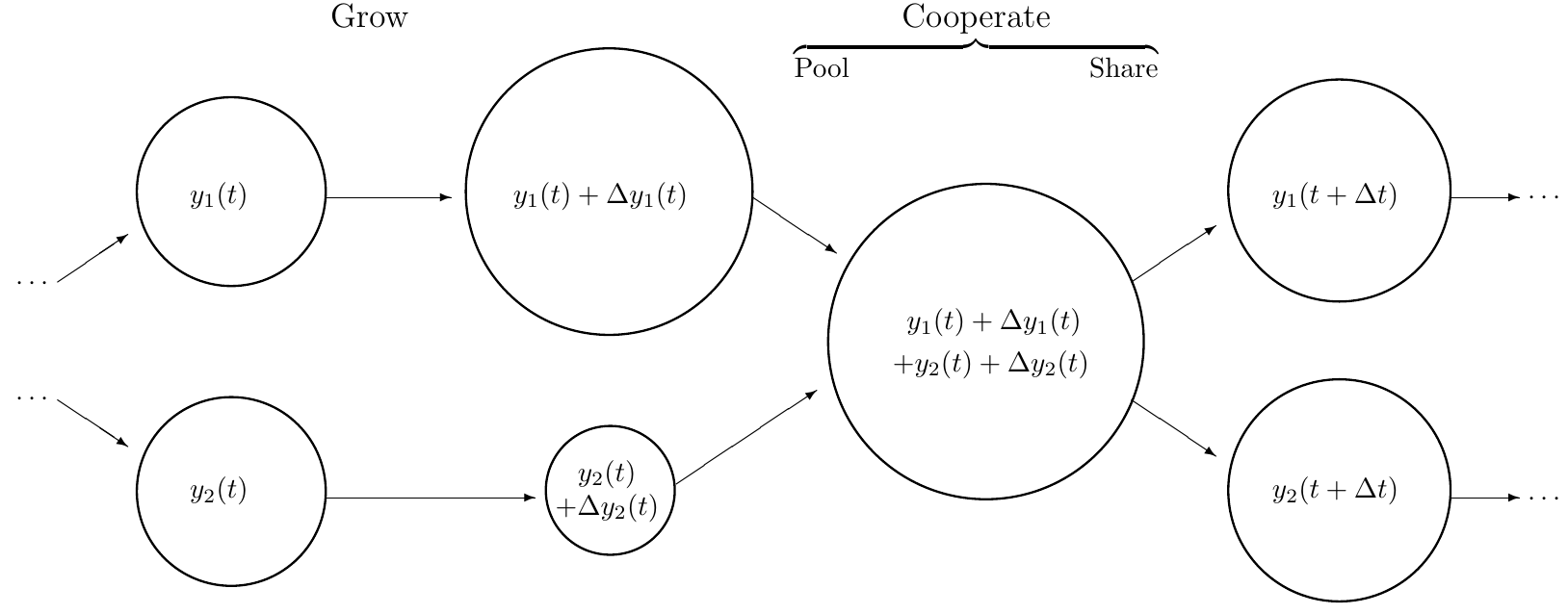}
\caption{Cooperation dynamics. Two cooperators start each time step with equal resources. Then they grow independently according to \eref{discrete_coop_grow}. Then they cooperate by pooling resources and sharing them equally, according to \eref{discrete_coop_coop}. Then the next time step begins.}
\flabel{dynamics}
\end{figure}
We imagine a mutation that hard-wires cooperation into the entities. Think, for example, of a mutation that causes cells to agglomerate by sharing their membranes. Previously independent entities with biomasses $x_i(t)$ now find themselves cooperating. We label their biomasses $y_i(t)$ to distinguish them from the equivalent non-cooperators. For simplicity we assume equal sharing, $y_i(t) = \yn{N}(t)$ for all $i$, where $\yn{N}(t)$ denotes the per-entity biomass for $N$ equal cooperators. 

In the discrete-time picture, each time step involves a two-phase process. First there is a growth phase, analogous to \eref{discrete_nonc_grow}, in which each cooperator increases its resources by
 \be
 \D y_i(t) = y_i(t)\left(\mu\D t + \sigma\xi_i\sqrt{\D t}\right).
 \elabel{discrete_coop_grow}
 \ee
This is followed by a cooperation phase, replacing \eref{discrete_nonc_coop}, in which resources are pooled and shared equally among the cooperators,
 \be
y_i(t+\D t) = \frac{1}{N}\sum_{i=1}^N\left(y_i(t) + \D y_i(t)\right) = y_i(t) + \frac{1}{N}\sum_{i=1}^N \D y_i(t),
 \elabel{discrete_coop_coop}
 \ee
where the second equality follows from the equality of the $y_i(t)$. This is equivalent to equal sharing of the total of the individual fluctuations,
\be
\D\yn{N}(t) = \frac{1}{N}\sum_{i=1}^N \D y_i(t).
\elabel{discrete_coop_change}
\ee

Cooperation has no direct cost in this protocol. In reality, pooling and sharing often require a coordinating mechanism. For example, large organisms have circulatory systems to redistribute nutrients, and human societies have administrative systems to redistribute resources. It is possible for such mechanisms to have costs that make otherwise beneficial cooperation unviable. We do not expect our model to describe well situations where such effects are important. Equally, we ascribe no direct benefit to cooperation. In our basic setup, costs and benefits emerge as the effects of cooperation on time-average growth rates.

Substituting \eref{discrete_coop_grow} into \eref{discrete_coop_change} yields the dynamic followed by the biomasses of each cooperator,
 \be
\Delta \yn{N}(t) = \yn{N}(t) \left(\mu \Delta t + \frac{\sigma}{N}\sum_{i=1}^N\xi_i \sqrt{\Delta t}\right).
 \elabel{discrete_cooperate}
 \ee
Sums of independent normal variates are normal. We define a standard normal variate,
 \be
 \xin{N} \equiv \frac{1}{\sqrt{N}} \sum_{i=1}^N \xi_i \sim \mathcal{N}(0,1),
 \elabel{define_xin_indept}
 \ee
which allows us to rewrite \eref{discrete_cooperate} as
\be
\Delta \yn{N}(t) = \yn{N}(t) \left(\mu \Delta t + \frac{\sigma}{\sqrt{N}} \, \xin{N} \sqrt{\Delta t}\right).
 \elabel{discrete_cooperate_combined}
\ee
Thus the net effect of $N$ individual fluctuations with pooling and equal sharing is a single equivalent fluctuation, whose amplitude is $1/\sqrt{N}$ times that of the individual fluctuations.  Substituting into \eref{discrete_cooperate} and letting the time increment $\Delta t \to 0$, we recover a stochastic differential equation of the same form as \eref{motion}, but with the volatility reduced from $\sigma$ to $\sigma/\sqrt{N}$:
 \begin{equation}
 d\yn{N} = \yn{N}\left(\mu dt +\frac{\sigma}{\sqrt{N}} \, d\Wn{N}\right).
 \elabel{gbm_coop}
 \end{equation}
 
\section{Analysis and solution of the cooperation puzzle}
\seclabel{analysis}
The expectation values of the biomasses of a non-cooperator, $\ave{x_1(t)}$, and the corresponding cooperator, $\ave{y_1(t)}$, with the same initial biomass, $x_1(0)$, are identical: $x_1(0)\exp(\mu t)$. From this perspective there is no incentive for cooperation. Moreover, immediately after the growth phase, \eref{discrete_coop_grow}, the better-off entity in a cooperating pair, say $y_1(0)>y_2(0)$, could increase its future expectation value from $[(y_1(0)+y_2(0))/2]\exp(\mu t)$ to $y_1(0)\exp(\mu t)$ by breaking the cooperative pact. Analyzing the growth of the expectation value gives no reason for cooperation to arise and, if it does arise, good reasons for it to end. From this perspective, cooperation looks fragile and its frequent observation in nature seems puzzling.

The solution of the puzzle comes from changing perspectives and considering the time-average growth rate instead of the ensemble-average growth rate. We know from \eref{gt} that non-cooperating entities grow at $\gtime(x_i)=\mu-\sigma^2/2$ over long time. The analogous growth rate for the cooperative dynamic, \eref{gbm_coop}, is found by changing $\sigma$ to $\sigma/\sqrt{N}$ in \eref{gt} to get
\be
\gtime\left(\yn{N}\right) = \mu - \frac{\sigma^2}{2N}.
\elabel{gt_coop}
\ee
For any non-zero volatility, cooperators have faster time-average growth rates than non-cooperators. The premium increases with the number of cooperators as $1-1/N$,
\be
\gtime\left(\yn{N}\right) - \gtime(x_i) = \frac{\sigma^2}{2}\left(1-\frac{1}{N}\right),
\ee
implying that larger cooperatives are favored over smaller ones.

From the perspective of our model, we see that cooperators will eventually dominate the environment and that cooperation will become ubiquitous. The effect is illustrated in \fref{cooperate} by direct simulation of \eref{discrete_nonc_grow}--\eref{discrete_nonc_coop} and \eref{discrete_cooperate}.
\begin{figure}
\includegraphics[width=\textwidth]{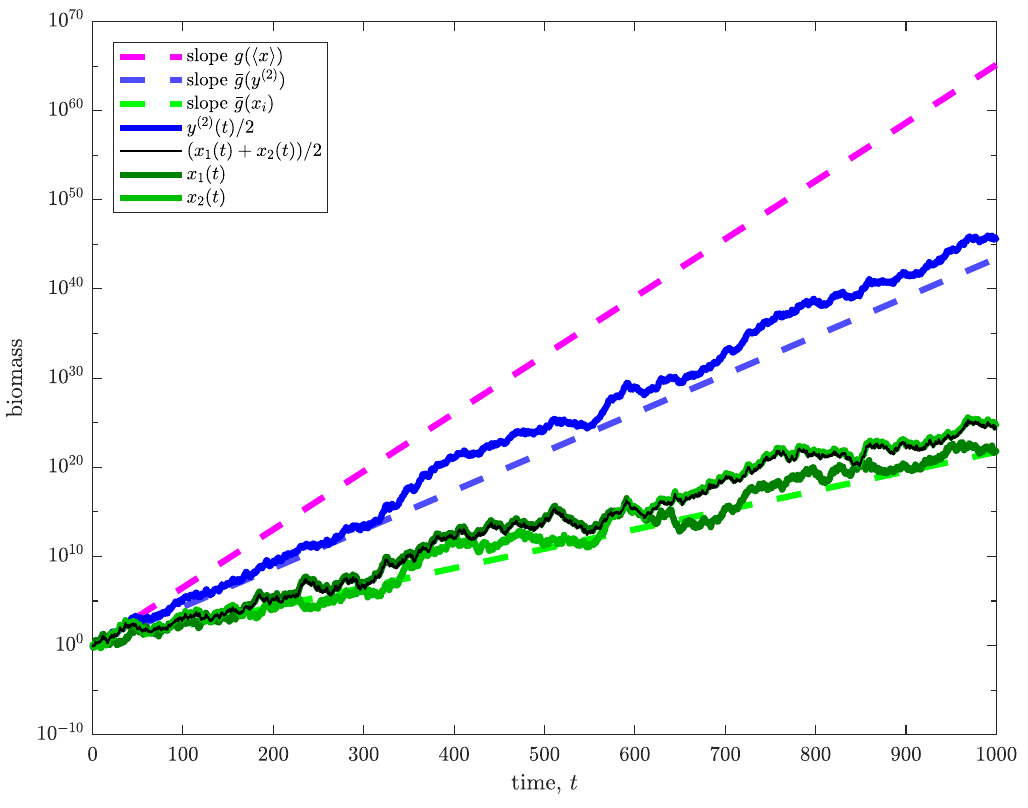}
\caption{Typical trajectories for two non-cooperating (green) entities and for the 
corresponding cooperating unit (blue) on a logarithmic vertical scale. 
Over time, the noise reduction for the cooperator leads to faster growth. Even without effects of specialization or the emergence of new function, cooperation pays in the long run. The thin black line shows the average of the non-cooperating entities, which is far inferior to the cooperating unit. In a very literal mathematical sense the whole, $y_1(t)+y_2(t)$, is more than the sum of its parts, $x_1(t)+x_2(t)$. The algebra of cooperation is not merely that of summation. The expectation-value growth rate (at which the pink dashed line grows) is the growth rate in the limit of infinitely many cooperators.}
\flabel{cooperate}
\end{figure}
Notice the nature of the Monte Carlo simulation in \fref{cooperate}. No ensemble
is constructed. Only individual trajectories are simulated and run for a time that is 
long enough for statistically significant features to emerge from the noise. This method teases out of the dynamics what happens over time.

Simulating an ensemble and averaging over members to remove noise is a different process which tells a different story. Features emerging from the noise when averaging over an ensemble do not, in general, also emerge over time. From \eref{gt_coop} we see that the time-average growth rate of a cooperator approaches the expectation-value growth rate of a non-cooperator (or, indeed, that of a member of a finite cooperative) only as the number of cooperators grows large,
\be
\lim_{N\to\infty}\gtime\left(\yn{N}\right) = \mu.
\elabel{gt_max}
\ee
The pink dashed line in \fref{cooperate} plots the temporal evolution of the expectation value of the biomass of a non-cooperator, which bears little resemblance to that of the biomasses of individual non-cooperators (green) and members of small cooperatives (blue).

\section{Generalizations of the model}
\seclabel{general}
\subsection{Idiosyncratic entities}
\seclabel{general-idio}
Real cooperatives have members of differing abilities as well as differing fortunes. The latter we model already as different realizations of the noise in \eref{motion}. The former we can treat by generalizing \eref{motion} so that the entities have idiosyncratic drifts and volatilities,
\begin{equation}
dx_i=x_i(\mu_i dt + \sigma_i dW_i),
\elabel{motion_idio}
\end{equation}
for $i=1\dots N$. Some entities will have higher individual time-average growth rates than others. This raises questions. Does it benefit leaders to share with laggards? When should a non-cooperator join a cooperating group? When should the group allow it?

Repeating the analysis of growth, pooling, and sharing yields a modified dynamic, 
 \bea
\D \yn{N}(t) &=& \yn{N}(t) \left(\frac{1}{N}\sum_{i=1}^N \mu_i \D t + \frac{1}{N}\sum_{i=1}^N\sigma_i\xi_i \sqrt{\D t}\right) \nonumber\\
&=& \yn{N}(t) \left(\mun{N} \D t + \sigman{N} \xin{N} \sqrt{\D t}\right).
 \elabel{discrete_cooperate_idio}
 \eea
$\xin{N}$ is a standard normal variate, as before, and
\be
\mun{N} \equiv \frac{1}{N}\sum_{i=1}^N\mu_i, \quad \sigman{N} \equiv \frac{1}{N} \sqrt{\sum_{i=1}^N\sigma_i^2}
\ee
are effective drift and volatility parameters. Therefore, the biomasses of the cooperators evolve according to
\be
 d\yn{N} = \yn{N}\left(\mun{N} dt +\sigman{N} d\Wn{N}\right),
 \elabel{gbm_coop_idio}
\ee
with time-average growth rate,
\be
\gtime\left(\yn{N}\right) = \mun{N} - \frac{(\sigman{N})^2}{2}
= \frac{1}{N}\sum_{i=1}^N\left(\mu_i-\frac{\sigma_i^2}{2N}\right).
\elabel{gt_coop_idio}
\ee
This is the sample mean of the time-average growth rates each entity would achieve by cooperating with like entities, \cf \eref{gt_coop}.

We can now answer the questions. It benefits entity $j$ to join the cooperative if $\gtime\left(\yn{N}\right)>\gtime(x_j)$, \ie if
\be
\frac{1}{N}\sum_{i=1}^N\left(\mu_i - \frac{\sigma_i^2}{2N}\right) > \mu_j-\frac{\sigma_j^2}{2}.
\elabel{coop_idio_condition}
\ee
Similarly, the cooperative benefits by admitting entity $j$ if
\be
\frac{1}{N}\sum_{i=1}^N\left(\mu_i- \frac{\sigma_i^2}{2N}\right) > \frac{1}{N-1}\sum_{\substack{i=1\\ i\neq j}}^N \left(\mu_i - \frac{\sigma_i^2}{2(N-1)}\right).
\ee
Consider two entities, where $\mu_1>\mu_2$ and $\sigma_1<\sigma_2$ so that $x_1$ grows faster over time than $x_2$. Rearranging \eref{coop_idio_condition} for $N=2$, we see that the fast grower should share with the slow grower if $\mu_1-3\sigma_1^2/4<\mu_2-\sigma_2^2/4$.

\subsection{Correlated fluctuations}
\seclabel{general-correl}
A second generalization concerns correlations. Fluctuations experienced by different entities are uncorrelated in our model: the $dW_i$ in \eref{motion} and, consequently, the $\xi_i$ in \eref{discrete_nonc_grow} onwards are independent random variables. In reality, cooperators are often spatially localized and experience similar environmental conditions. By allowing correlations between the $\xi_i$, our model can be adapted to describe such situations.

Suppose the $\xi_i\sim\mathcal{N}(0,1)$ realized in a given time step are jointly normal and cross-correlated such that $\ave{\xi_i\xi_j}=\rho_{ij}$. Assume for simplicity that
\be
\rho_{ij} =
\begin{cases}
1, & i=j,\\
\rho, & i\neq j,
\end{cases}
\ee
\ie that the fluctuations for all pairs of different entities have the same covariance, $\rho$, where $\lvert\rho\rvert\leq1$. The more general case of a covariance matrix with unequal off-diagonal elements is also tractable, but adds complexity without illumination.

The presence of cross-correlations alters the evaluation of the sum of the normal variates in \eref{discrete_cooperate}. We have now
\be
\sum_{i=1}^N \xi_i \sim \mathcal{N}(0, N + N(N-1)\rho).
\elabel{sum_correl}
\ee
Positive variance requires $\rho$ to be confined to $-1/(N-1)\leq\rho\leq1$. \Eref{sum_correl} suggests defining, analogously to \eref{define_xin_indept}, a standard normal variate,
\be
\xin{N} \equiv \frac{1}{\sqrt{N+N(N-1)\rho}}\sum_{i=1}^N\xi_i \sim \mathcal{N}(0,1),
\elabel{define_xin_correl}
\ee
such that the change in $\yn{N}$ can be written as
\be
\D\yn{N}(t) = \yn{N}(t) \left(\mu \D t + \sigma\sqrt{\frac{1+(N-1)\rho}{N}} \, \xin{N} \sqrt{\D t}\right),
\elabel{discrete_cooperate_combined_correl}
\ee
analogous to \eref{discrete_cooperate_combined} in the uncorrelated case.

Without correlations, cooperation reduces the amplitude of fluctuations from $\sigma$ to $\sigma/\sqrt{N}$. With them, the fluctuation amplitude becomes
\be
\sigmac \equiv \sigma\sqrt{\frac{1+(N-1)\rho}{N}}.
\elabel{sigma_correl}
\ee
The variation of $\sigmac$ with $\rho$ and $N$ delineates the main features of this generalized model. Firstly, as a consistency check, we note that $\sigmac\to\sigma/\sqrt{N}$ as $\rho\to0$ for fixed $N$, recovering the uncorrelated result in the appropriate limit.

For all $N>1$, we have $0\leq\sigmac\leq\sigma$, with $\sigmac=\sigma$ if and only if $\rho=1$. In other words, provided fluctuations are not perfectly correlated, a cooperation benefit always exists. This makes intuitive sense. With perfect correlation, all the $\xi_i$ are identical and pooling and sharing achieves nothing. The cooperative is equivalent to a giant individual following a single trajectory of \eref{motion}. As soon as some variation is introduced between the fluctuations of the entities, the cooperation mechanism can begin to mitigate the negative effects of fluctuations on growth.

Furthermore, we have $\sigmac\to\sigma\sqrt{\rho}$ as $N\to\infty$, with $0\leq\rho\leq1$ in this limit. The maximum time-average growth rate achievable by adding cooperators is, therefore,
\be
\lim_{N\to\infty}\gtime\left(\yn{N}\right) = \mu-\frac{\sigma^2\rho}{2},
\elabel{gt_max_correl}
\ee
\cf \eref{gt_max}. This cannot exceed $\mu$ (since $\rho$ is non-negative) and decreases as $\rho$ increases. Again, this is consistent with intuition: as fluctuations become more correlated, the variation between them diminishes and the scope for beneficial cooperation narrows. In our setup, cooperation relies on diversity in individual outcomes.

\subsection{Partial cooperation\seclabel{general-partial}}
Another generalization of our model is to partial cooperation. Here entities pool and share only a fraction of their resources, resembling, for example, taxation and redistribution in human societies. We discuss this in a separate manuscript \citep{BermanPetersAdamou2017}.

\section{Discussion}
\seclabel{discussion}
Our model assumes nothing more than that evolutionary processes are multiplicative and noisy. In this context, the evolutionary advantage of cooperation arises from the nonlinear dependence of growth rates on temporal fluctuations. By reducing the amplitude of fluctuations, pooling and sharing increase the time-average growth rate of cooperating entities. This paints a picture of cooperation driven by self-interest, not altruism, with cooperators outgrowing similar non-cooperators.

Where our model describes well the growth of things in nature, it predicts that cooperation will be prevalent. As an attractor for many processes whose relative changes are independent random variables, we expect the model to resemble many real examples of self-reproduction.

In reality, many effects contribute to the formation of cooperative structure. Members of large cooperatives can coordinate their actions to develop emergent functions, such as the ability to swim towards nutrients. Cells in multicellular organisms differentiate and specialize in particular tasks. Human cooperation works analogously, with firms and individuals becoming proficient in different roles.

By finding a cooperative benefit under minimal assumptions, our analysis may shed light on cases of cooperation without such effects. In particular, it may explain the transition from single cells to bicellular organisms, too small and simple to benefit from new function or specialization.

Classical treatments of cooperation are predicated on the existence of a net fitness gain in a donor-recipient interaction. A universal mechanism by which this benefit arises would strengthen their theoretical foundations. In our model of noisy self-reproduction, we identify the increase in time-average growth rate achieved through cooperation as this net fitness gain. This is consistent with the concept of geometric mean fitness in the biological literature.

The impact of risk reduction on long-time growth suggests that risk management has a rarely recognized significance. Fluctuation reduction, or good risk management, does not merely reduce the likelihood of disaster or the size of up and down swings. It also improves the long-time performance of the structure whose risks are being managed. While the effect of reducing fluctuations depends on the specific setup, it is tantalizing to see that the simple and universal setting of multiplicative growth favors structure in the form of large cooperative units.

The insight that time averages may not be identical to expectation values
was only reached in the development of statistical mechanics in the
$19^\text{th}$ century, where the physical nature of expectation
values was identified as the mean over an ensemble of non-interacting systems.
The development of ergodic theory in the $20^\text{th}$ and $21^\text{st}$ centuries provided the  concepts that reveal the physically relevant aspect of stochastic processes like \eref{motion}. Today we have at our disposal mathematical tools that allow us to understand nonlinear dynamical effects, such as described here. Applying these tools suggests that our natural tendency to cooperate -- manifested in our gut feeling and moral sentiment -- is in harmony with a careful formal analysis of the issues involved. 

\section*{Acknowledgements}
We acknowledge D.\ Krakauer for helpful comments and Baillie Gifford for supporting the ergodicity economics program at the London Mathematical Laboratory.

\bibliographystyle{humannat}
\bibliography{./bibliography}   
\end{document}